\documentclass[fleqn,10pt, twocolumn]{wlscirep}
\usepackage[utf8]{inputenc}
\usepackage[T1]{fontenc}
\usepackage{bm}
\usepackage{fancybox}
\usepackage[font=small,labelfont=bf,justification=justified]{caption}
\usepackage{newfloat}
\usepackage{txfonts}
\DeclareFloatingEnvironment[fileext=los, listname={List of Boxes}, name=Box, placement=tbhp, within=none]{boxfloat}

\title{Benchmarking quantum computers}
\author[1,*]{Timothy Proctor}
\author[1]{Kevin Young}
\author[2]{Andrew D. Baczewski}
\author[3]{Robin Blume-Kohout}
\affil[1]{Quantum Performance Laboratory, Sandia National Laboratories, Livermore, CA 94550, USA}
\affil[2]{Quantum Algorithms \& Applications Collaboratory, Sandia National Laboratories, Albuquerque, NM 87185, USA}
\affil[3]{Quantum Performance Laboratory, Sandia National Laboratories,  Albuquerque, NM 87185, USA}
\affil[*]{e-mail: tjproct@sandia.gov}

\begin{abstract}
The rapid pace of development in quantum computing technology has sparked a proliferation of benchmarks for assessing the performance of quantum computing hardware and software. Good benchmarks empower scientists, engineers, programmers, and users to understand a computing system's power, but bad benchmarks can misdirect research and inhibit progress. In this Perspective, we survey the science of quantum computer benchmarking. We discuss the role of benchmarks and benchmarking, and how good benchmarks can drive and measure progress towards the long-term goal of useful quantum computations, i.e., “quantum utility”. We explain how different kinds of benchmark quantify the performance of different parts of a quantum computer, survey existing benchmarks, examine recent trends in benchmarking, and highlight important open research questions in this field. 
\end{abstract}
\begin{document}
\flushbottom
\maketitle
\thispagestyle{empty}

\begin{figure}[t!]
    \centering
      \includegraphics[width=8.5cm]{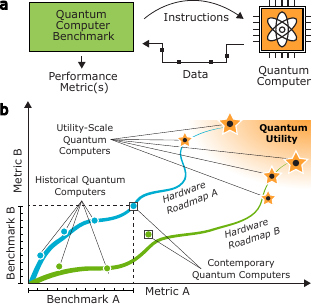}
    \caption{\textbf{Quantum computer benchmarks.} \textbf{a.} Quantum computer benchmarks are methods that are run on quantum computing systems, or on some of their subsystems (qubits, compilers, etc), to measure performance. Each benchmark measures one or more metrics of performance, such as the error rates of a system's quantum gates. \textbf{b.} Benchmarks enable comparing a quantum computer's performance to other contemporary, historical, or hypothetical systems, e.g., hypothetical systems on a particular roadmap to quantum utility.
    \label{fig:schematic}}
\end{figure}

\noindent 
In quantum computing, \emph{benchmarking} means measuring the performance of quantum computing systems, and \emph{benchmarks} are methods that are used to measure performance (Fig.~\ref{fig:schematic}a). The rapid advance of quantum computing hardware over the past decade has spawned a bewildering variety of concepts and techniques within the fledgling science of quantum computer benchmarking---a subfield of QCVV (quantum characterization, verification, and validation) \cite{Hashim2024-om}. In 2014, cutting edge “quantum computers” were physics experiments on a few qubits \cite{Barends2014-ya}. Today’s quantum computers routinely deploy 20-400 qubits \cite{Arute2019-mk, Bluvstein2023-dp, Kim2023-si, Moses2023-do, Chen2023-la} and can outperform classical computers on specially designed tasks \cite{Arute2019-mk}. But despite this progress, quantum computers are still in their infancy. Our field’s goal of solving problems of practical significance---often called “quantum utility”---remains elusive, and many scientists predict that achieving quantum utility will require years of progress and major technological advances \cite{gidney2021factor, lee2021even, Rubin2024-gc, Childs2018-ms}. Benchmarking and benchmarks can measure and guide progress toward that goal.  

Benchmarks for classical computers, such as the LINPACK \cite{Dongarra2003-fw} and SPEC \cite{SPEC} benchmarks, measure proxies for computational utility. Once quantum computers become capable of computational utility, many quantum computer benchmarks will too. But today, and in the near term, quantum computer benchmarks should instead measure, guide, and incentivize progress towards utility (Fig.~\ref{fig:schematic}b). To solve practical problems that classical computers cannot, it is expected that quantum systems will need to grow to millions of physical qubits \cite{Childs2018-ms,gidney2021factor, Rubin2024-gc, lee2021even}. But more qubits is not enough. Achieving quantum utility requires reducing and mitigating hardware errors, which are the imperfections in qubits and quantum logic gates (simply ``gates'' hereafter) that cause large quantum programs to fail (see Box 1). The importance of other limiting factors, like speed or power consumption, pales in comparison. Quantum logic operations fail far more often (today, typically around 0.01\%-1\% of the time \cite{Arute2019-mk, Bluvstein2023-dp, Kim2023-si, Moses2023-do, Chen2023-la}) than classical instructions (around $10^{-23}\%$ of the time \cite{Ziegler1979-st}), and correcting or mitigating quantum errors is startlingly hard \cite{Campbell2017-tw, Eastin2009-jm, litinski2019magic, Bluvstein2023-dp, Krinner2022-tp, Google_Quantum_AI2023-yr, Gupta2024-pr}. Because errors will dominate the performance and design of quantum computers for the foreseeable future, most quantum computer benchmarks measure the impacts of those errors.

Many of the earliest and most widely-used quantum computer benchmarks, such as randomized benchmarking (RB) \cite{Emerson2005-fd, Emerson2007-am, Knill2008-jf, Magesan2011-hc, Proctor2019-gf, Hines2023-tz, McKay2023-bx, Proctor2022-yl, Hines2023-vq, Combes2017-kr, Helsen2019-cp, Helsen2022-yp,  Magesan2012-dz}, directly measure the rates of errors in a quantum computer's gates. But quantum computers are not just arrays of qubits. Instead, a phalanx of powerful classical computing subcomponents---controllers, compilers, routers, and schedulers---manage and control the qubits \cite{Corcoles2020-vn}. Increasingly, quantum computations will be performed not directly on physical qubits, but instead on higher-performing \emph{logical qubits} encoded within many physical qubits running a quantum error correction (QEC) algorithm \cite{Campbell2017-tw}. These complexities have motivated an ever-increasing array of incomparable and complementary benchmarks that test different subsystems in a quantum computer and/or measure different performance metrics. Understanding and leveraging a benchmark requires knowing what systems it tests, what it measures about those subsystems' performance, and why that metric of performance matters.
 
In this perspective, we survey the science of quantum computer benchmarking and we highlight important open research questions in this field. Our focus is on benchmarking rather than the broader field of QCVV, and we present a forward-looking and opinionated perspective that complements recent reviews and tutorials on QCVV \cite{Hashim2024-om, Eisert2020-an, Kliesch2021-ea} and Carrasco \emph{et al.}'s perspective on verifying results from quantum devices \cite{Carrasco2021-ep}. Throughout, we limit our scope to the benchmarking of gate-model quantum computers built from qubits (the primary quantum computing paradigm), e.g., we do not discuss benchmarks relevant only to quantum annealers or analog quantum simulators. We presume that the goal of quantum computing R\&D is to achieve and then advance computational utility, and we measure the relative value of benchmarks by how useful and appropriate they are to that goal.

We present our perspective in three sections. In the first section, we discuss the origins of quantum computer benchmarking, current trends in the field, and the roles that benchmarks play within the quantum computing community. Having summarized the current state of play, we then discuss what properties make a benchmark broadly useful. In the second section, we provide a framework within which benchmarks can be categorized. We use this framework to analyze many of the most important and popular benchmarks and highlight specific gaps in the benchmarking toolbox. In the final section, we turn to what is, in our view, the largest open question in benchmarking: how can our field create and deploy benchmarks that reliably quantify progress towards quantum utility? We discuss how these benchmarks might be created, presenting our view on the ingredients needed from across the quantum computing community.

\begin{boxfloat}[t!]
\begin{center}
\fbox{%
\hspace{-0.05cm}
\begin{minipage}{.47\textwidth}
\small{
\vspace{0.05cm}
\textbf{Box 1. Errors in quantum computers.} Gate-model quantum computers implement computations using quantum circuits, which are sequences of gates \cite{Nielsen2012-vg} [see example below, containing the Hadamard (H), phase (P) and CNOT gates (connected circles)]. When circuits are run on quantum computing hardware, errors can occur (red star, below). Errors cause quantum computations to fail by corrupting the outputs of quantum circuits, and most benchmarks quantify the magnitude and/or impact of these errors.
\vspace{0.05cm}

\includegraphics[width=8.3cm]{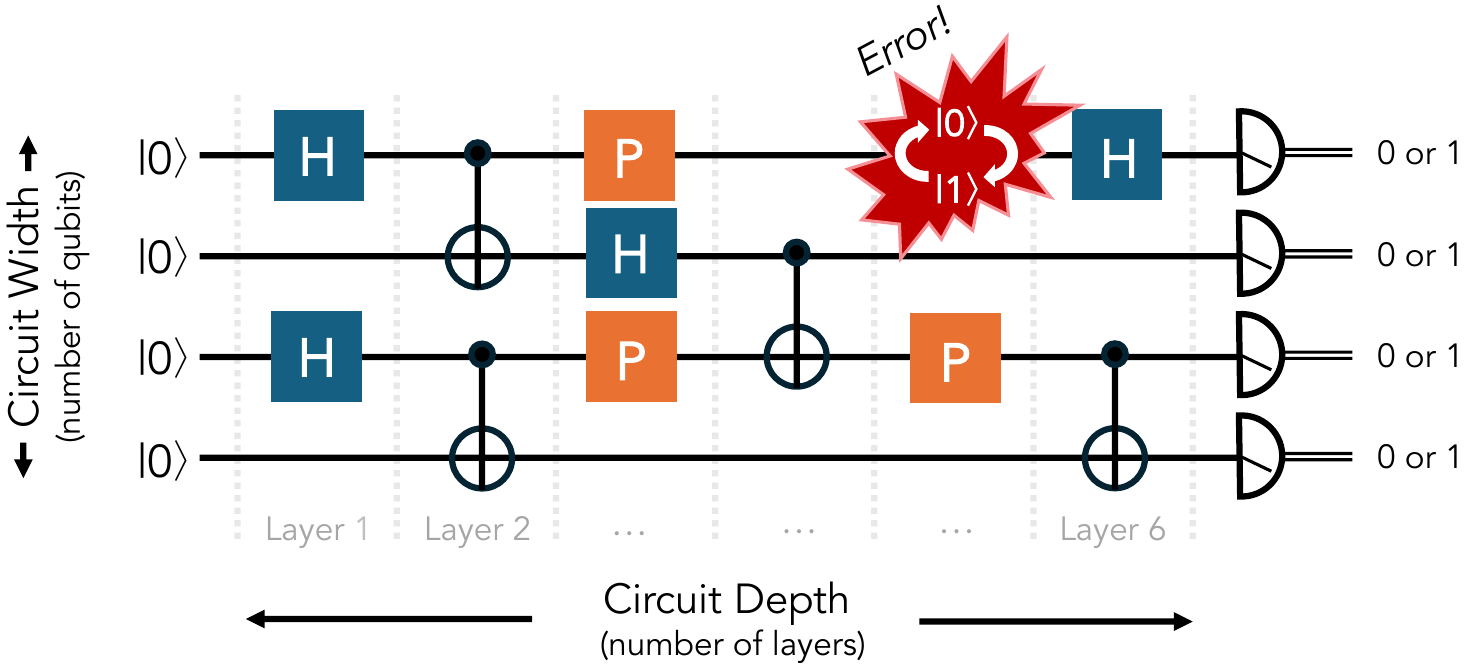}

\hspace{0.3cm}
A simple and popular model for errors in quantum circuits is that a random bit (and/or phase) flip occurs with some probability $\epsilon$ at each location in a circuit. So the probability of at least one error in a circuit of width $w$ and depth $d$ is $1-(1-\epsilon)^{wd} \lesssim wd \epsilon$, implying that a computation will likely succeed only if $\epsilon \ll 1/(wd)$. In this simple model, a quantum computer's errors are described by a single error rate ($\epsilon$). But real quantum computers experience diverse and \emph{a priori} unknown errors, and each kind of error has different effects \cite{Proctor2021-wt, Blume-Kohout2022-ln, Murphy2019-bp}. Important examples of error types include coherent (i.e., systematic) and stochastic Pauli errors \cite{Blume-Kohout2022-ln}, crosstalk \cite{Sarovar2020-pz, Gambetta2012-zd}, and slow drifts in a system's parameters \cite{Proctor2020-iz}. The complexities of the errors that occur in real quantum computers makes diverse, complementary benchmarks and metrics necessary.
\vspace{0.05cm}
\label{box:errors}}\end{minipage}
\hspace{-0.05cm}}
\end{center}
\end{boxfloat}

\section*{Quantum computer benchmarking}
The field of quantum computer benchmarking evolved out of quantum tomography and error characterization \cite{Chuang1997-dj, Gale1968-hp, Hradil1997-tg, Poyatos1997-jv,  Nielsen2021-nu}. These are techniques for reconstructing quantum states and learning detailed error models for qubits and quantum computers. The earliest ``quantum computers'' were physics experiments on one or two qubits, and physicists used characterization methods to learn about how their systems behaved \cite{Monroe1995-tf, Chuang1998-am}. Early benchmarks (e.g., RB \cite{Emerson2005-fd,  Knill2008-jf, Emerson2007-am}) emerged to provide simpler, more concise assessments of one or two qubit operations.  As $n$-qubit systems with $n\geq 5$ emerged, full tomographic characterization of all $n$ qubits grew rapidly infeasible.  It became both convenient and necessary to summarize the ``holistic'' performance of prototype quantum computers using a handful of metrics measured by benchmarks \cite{Cross2019-ku, Linke2017-mr, Blume-Kohout2020-de, Boixo2018-kp, Erhard2019-wk, Proctor2019-gf}. 

A quantum computer benchmark (or just ``benchmark'' hereafter) is defined by (i) a set of computational tasks together with a procedure for performing those tasks on a quantum computer (or one or more of its components), and (ii) a method for computing one or more performance metrics from the (classical) data produced by running those tasks (Fig.~\ref{fig:schematic}a). This definition for a benchmark encompasses most (but not all) uses of the term in quantum computer science, and it defines our scope. Benchmarks that measure metrics that are not direct computational properties of quantum computing systems (e.g., qubit coherence times, or heating rates of ion traps) are outside our scope, including benchmarks that measure a qubit's $T_1$ and $T_2$ times \cite{Hashim2024-om}, which are widely-used metrics for qubit quality.

For most benchmarks, the tasks comprise one or more quantum circuits, and for many benchmarks the output is a single, scalar metric of performance, e.g., quantum volume \cite{Cross2019-ku}. However, a single benchmark can compute multiple metrics (e.g., a success rate and a run time), or non-scalar quantities such as capability regions\cite{Proctor2021-wt} (see Fig.~\ref{fig:progress}). In our view, no single benchmark or performance metric can capture all important aspects of a quantum computer's performance, as is accepted with classical computers.

\subsection*{From low-level to high-level benchmarks}
\begin{figure}[t!]
    \centering
    \includegraphics[width=8.5cm]{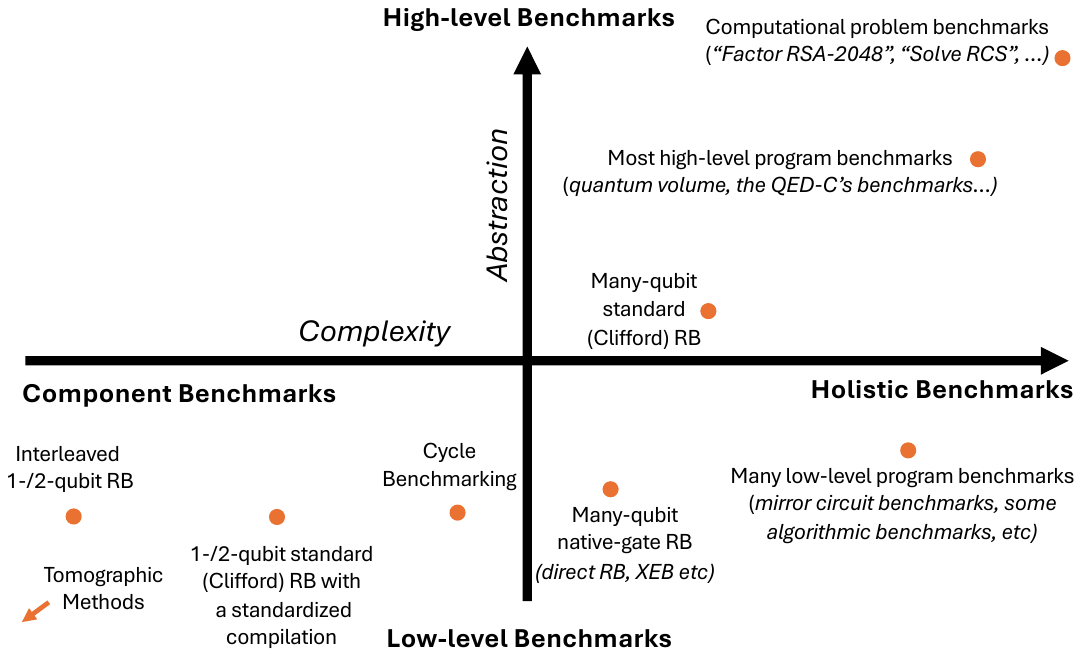}
    \caption{\textbf{Kinds of benchmark}. Benchmarks vary widely in the \emph{abstraction} level of their tasks, ranging from executing specific low-level quantum circuits to solving a computational problem, and by the \emph{complexity} of the object whose performance they measure, ranging from individual gates to entire computing systems. The abstraction and complexity of a selection of important benchmarks and benchmark families are shown here. We also show tomographic (characterization) methods (e.g., state, process, and gate set tomography), which are not benchmarks, for comparison.
    \label{fig:kinds-of-benchmark}}
\end{figure}

The field of quantum computer benchmarking began in the early 2000s with the initial development of RB \cite{Emerson2005-fd, Emerson2007-am, Knill2008-jf} as well as pioneering experiments on 2-12 qubits using quantum algorithms for solving tiny problems or creating canonical states as informal benchmarks \cite{Gulde2003-wq, Negrevergne2006-mz, Lanyon2007-rk, DiCarlo2009-jw, Knill2000-uk}. The purpose of all these benchmarks was to directly quantify the rates of errors associated with qubits and gates, and their impact on quantum circuits (Box 1 reviews errors and circuits). For example, RB measures the error rates of gates. These earliest methods are \emph{low-level} benchmarks (see Fig.~\ref{fig:kinds-of-benchmark}) that isolate qubit and gate performance from classical components (e.g., compilers) within an integrated quantum computing system. Since 2005, many low-level benchmarks that measure gate and circuit error rates have been (and continue to be) developed. They complement these earliest methods or address their limitations, and we discuss several of them in the next section.

In the last decade, benchmarking research has increasingly focused on \emph{high-level} benchmarks (see Fig.~\ref{fig:kinds-of-benchmark}) that quantify the overall performance of integrated (a.k.a.~full-stack) quantum computers. The most well-known high-level benchmark is IBM's quantum volume benchmark \cite{Cross2019-ku}. High-level benchmarks still predominantly quantify the impact of errors in gates and qubits, but they do so by measuring their effect within the context of a complete quantum computing ``stack'' (see Fig.~\ref{fig:stack}) that potentially includes a lot of classical computing subsystems. High-level benchmarks can quantify the impact of errors from the perspective of a quantum computer user executing high-level quantum programs. But to do so, they mix together many different aspects of performance \cite{Corcoles2020-vn, Jurcevic2021-dz}, including the performance of purely classical subsystems, that may not all be relevant to the pursuit of quantum utility.

The trend towards high-level benchmarks and metrics over the last decade has been driven by hopes for near-term quantum utility with \emph{noisy intermediate-scale quantum} (NISQ) computing \cite{Preskill2018-jz, Bharti2022-rz, Chen2023-rx} (see Box 2), by the emergence of commercial quantum computing systems with tightly integrated classical software stacks \cite{Corcoles2020-vn}, and by the needs of prospective quantum computer users with limited quantum computing expertise. In the last few years, this trend has culminated with the development of many benchmarking suites centered around quantum algorithms and applications \cite{Chen2022-dm, Linke2017-mr, Wright2019-zj, Tomesh2022-nu, Murali2019-my, Donkers2022-wt, Finzgar2022-aa, Mills2020-zh, Lubinski2023-zy, Lubinski2024-ci, Lubinski2023-mr, Chen2023-la, Benedetti2019-pp, Li2020-ry, Quetschlich2023-bg, Dong2021-gj, Martiel2021-vp, Van_der_Schoot2022-gv,Van_der_Schoot2023-vo, Cornelissen2021-yt, Georgopoulos2021-hh, Dong2022-ga}. Although these trends are suggestive of a technology that is at or near utility, achieving utility seems like it will require major technological advances, notably ultra-reliable logical qubits (see Box 2). As long as quantum computing is in its infancy, our opinion is that high-level benchmarks---including those based on algorithms and applications---should be designed, used, and interpreted with caution.

\subsection*{The diverse roles and influences of benchmarks}
In principle, benchmarks are passive tools for measuring performance metrics, used to localize devices in ``performance metric space'' (see Fig.~\ref{fig:schematic}b). But benchmarks also exert sociological forces. They influence decisions both small (e.g., gate calibrations) and large (e.g., funding allocations), and drive resources towards optimizing performance on those benchmarks.  Benchmarks are used to compare competing quantum computing technologies and make judgements (sometimes implicit) about the current and future merits of each technology \cite{Linke2017-mr, Moses2023-do, Lubinski2023-zy, Tomesh2022-nu, Van_der_Schoot2022-gv, Martiel2021-vp, Donkers2022-wt}. They therefore influence the design of future quantum computing systems. Benchmarks are also used to optimize individual quantum computing systems, by choosing system settings (e.g., compilation parameters or gate pulses) that maximize performance on a benchmark \cite{Jurcevic2021-dz, Kelly2014-my, Rol2017-yw}, sometimes with automated routines \cite{Kelly2014-my, Rol2017-yw, Egger2014-gq}. 

Using benchmarks to inform a quantum computing system’s design and settings can easily lead to unproductive choices, because the chosen benchmarks might not summarize all important aspects of performance. Fundamental design decisions primarily aimed at maximizing performance on any existing high-level benchmarks would be particularly shortsighted in our view, because no existing high-level benchmarks have been shown to quantify progress towards quantum utility. For example, quantum volume favors systems with high qubit connectivity \cite{Corcoles2020-vn, Jurcevic2021-dz, Pino2021-zk, Moses2023-do} but higher connectivity will not necessarily enable better logical qubits. Optimizing an existing system’s settings to maximize its performance on benchmarks is less consequential, but can also result in poor performance unless the benchmarks are matched to that system's intended uses. For example, using (only) one- and two-qubit RB to calibrate the gates of a system \cite{Kelly2014-my, Rol2017-yw, Egger2014-gq} intended to run many-qubit circuits (e.g., NISQ algorithms or QEC) is unhelpful if that system has significant crosstalk errors \cite{Sarovar2020-pz, Gambetta2012-zd, Proctor2022-yl, Harper2023-pv} that could be reduced by calibrations, because these benchmarks are largely insensitive to those errors \cite{McKay2019-ca, Proctor2022-yl}. 

Widely adopted benchmarks and metrics that become \emph{de facto} or official standards are particularly influential. Today, quantum volume is the most popular metric for summarizing an integrated quantum computer's overall performance \cite{Pino2021-zk, Jurcevic2021-dz}. A quantum computer's total number of qubits and its average gate error rate (measured using RB) are the most popular metrics for summarizing its low-level performance. These metrics are informative, but any small set of metrics has limitations. These metrics alone do not reliably quantify progress towards quantum utility (see later), so our opinion is that caution is needed when using them to compare quantum computing systems.

\subsection*{Desirable properties of benchmarks}
The broad influence of benchmarks makes it critically important to design, adopt, and carefully use ``good'' benchmarks. In our view, a benchmark is good if it aids progress towards quantum computational utility (and eventually towards increasing utility) when used correctly, and if it is difficult to misuse. We do not know any algorithm for guaranteeing that a benchmark will satisfy these desiderata. But we can identify desirable properties that, in our opinion and experience, tend to make benchmarks useful and discourage misuse. Benchmarks that do not satisfy these properties are less likely to aid progress toward utility, and are more easy to misuse and misinterpret.
\begin{enumerate}
\item \emph{Well-motivated}. A benchmark should measure well-motivated metrics of performance.
\item	\emph{Well-defined}. A benchmark should have an unambiguous procedure, i.e., any unspecified steps in that procedure should be intentional configurable parameters (e.g., some benchmarks allow creative compilation---see Box 3).
\item	\emph{Implementation-robust}. It should not be possible to exploit (``game'') a benchmark's configurable parameters to obtain misleading results.
\item	\emph{System-robust}. A benchmark's results should not be corrupted when the tested quantum computer experiences \emph{a priori} unknown (small) errors.
\item	\emph{Efficient}. A benchmark should use a reasonable amount of all resources (e.g., quantum and classical computer time).
\item	\emph{Technology independent}. A benchmark should specialize to particular technologies or architectures only inasmuch as its metrics are only relevant in those contexts.
\end{enumerate}

\begin{boxfloat}[t!]
\begin{center}
\fbox{%
\hspace{-0.05cm}
\begin{minipage}{.47\textwidth}
\small{
\vspace{0.05cm}

\textbf{Box 2. NISQ and fault-tolerant quantum computing.} Quantum algorithms can be executed directly on physical qubits, which are not inherently fault tolerant. This approach, using systems akin to those available today, is referred to as \emph{noisy intermediate-scale quantum} (NISQ) computing \cite{Preskill2018-jz}. NISQ computers possess a moderate number of qubits, ranging from approximately 20 to 1000, and exhibit moderate error rates, between about $1\%$ and $0.1\%$. It is hoped that NISQ computers will demonstrate quantum utility through the use of heuristic algorithms specifically designed for such systems, commonly known as NISQ algorithms \cite{Preskill2018-jz, Bharti2022-rz, Chen2023-rx}. However, achieving utility-scale instances of quantum algorithms that exhibit a clear quantum advantage, such as Shor's algorithm \cite{Shor1994-zh}, is expected to require billions of gates \cite{gidney2021factor, lee2021even, Rubin2024-gc, Childs2018-ms}. Consequently, executing these algorithms successfully necessitates error rates around $10^{-9}$ or lower, a target currently considered infeasible for physical qubits.

\hspace{0.3cm}
Quantum computations can be rendered tolerant to errors by employing redundantly encoded \emph{logical qubits}, protected using fault-tolerant quantum error correction (QEC) techniques \cite{Campbell2017-tw, Bluvstein2023-dp, Krinner2022-tp, Google_Quantum_AI2023-yr, Gupta2024-pr}.  According to fault tolerance theory, these logical qubits can achieve significantly lower error rates than their physical counterparts, provided the errors in the physical qubits are sufficiently rare and exhibit desirable characteristics (e.g., are adequately uncorrelated) \cite{Aharonov2008-kv}. In fault-tolerant quantum computing, physical qubits undergo repeated cycles of QEC, which differ significantly from the operations used in NISQ algorithms. This distinction necessitates different benchmarks and metrics for evaluating prototype NISQ and fault-tolerant quantum computers.

\hspace{0.3cm}
Quantum algorithms are compiled very differently for NISQ and fault-tolerant architectures, and typically require many more gates (e.g., $50\times$ more) for fault-tolerant execution. While most physical qubits can be directly manipulated using a continuous, universal set of one- and two-qubit gates \cite{Nielsen2012-vg}, logical qubits are restricted to a discrete and non-universal set of ``easy'' gates, as per the Eastin-Knill theorem \cite{Eastin2009-jm}. Achieving universal fault-tolerant quantum computation requires the use of ``hard'' gates, which require additional resources for implementation (e.g., magic state distillation \cite{litinski2019magic}). This distinction has significant implications for the development of benchmarks for quantum computers.
\vspace{0.05cm}
\label{box:errors}}\end{minipage}
\hspace{-0.05cm}}
\end{center}
\end{boxfloat}

These properties describe ideals against which we think it is useful to judge benchmarks. But they are not precise or binary properties, and not every useful benchmark scores highly on all of them. For example, informal benchmarks are often created to demonstrate experimental advances and these benchmarks need only serve the scientific purposes at hand (so, e.g., implementation robustness is irrelevant). However, a benchmark warrants trust from the entire quantum computing community only if it approximates these ideals. In our opinion, few (and probably none) of today's high-level benchmarks achieve this standard.

Benchmarks must be implementation- and system-robust if they are to enable reliable comparisons between different technologies, or positively influence engineering decisions. Benchmarks are ideally designed to measure independently defined and well-motivated metrics (e.g., gate fidelity), and such benchmarks are robust if they accurately measure the intended metrics under broad conditions. However, proving that a benchmark is robust is often difficult, and is the central task in much of the theory of benchmarks. The accepted standard is to prove that a benchmark accurately measures its metric for any system whose qubits experience small Markovian errors \cite{Blume-Kohout2022-ln}, as exemplified by the theory of RB \cite{Helsen2022-yp, Merkel2021-ux, Proctor2017-wc, Wallman2018-yc}. However, qubits often experience significant non-Markovian errors, and those errors can often corrupt existing benchmarks (e.g., leakage can corrupt RB \cite{Wood2018-wf}). Theories and benchmarks that address more general errors are beginning to be developed \cite{Figueroa-Romero2021-xe, Figueroa-Romero2024-uk,Wood2018-wf, Proctor2020-iz} but remain a need for the field.

Many high-level benchmarks measure self-defined metrics (e.g., quantum volume is defined as the quantity measured by its eponymous benchmark \cite{Cross2019-ku}) and this complicates assessing those benchmarks' robustness. Such benchmarks are always intended to measure proxies for some independent (if perhaps imprecisely defined) aspect of performance. For example, high-level benchmarks based on one or more algorithms are typically meant to quantify how well a system can run ``interesting'' instances of those algorithms (or how far the system is from running those interesting instances), but they do not necessarily do so. The intended interpretation of a benchmark implies an imprecise notion of robustness that can be assessed. Arguably, the quantum volume benchmark is robust \cite{Corcoles2020-vn, Jurcevic2021-dz}, whereas many other existing high-level benchmarks are not. However, the inherent imprecision and subjectivity in assessing the merits of benchmarks with self-defined metrics is unfortunate. The field would benefit from the creation of high-level benchmarks that robustly measure well-motivated and independent performance metrics.

Benchmark efficiency is necessary to make a benchmark practical, but it is surprisingly challenging to design efficient benchmarks. Many benchmarks are intended to be run on $n$ qubits for any user-specified $n$, and such a benchmark is formally efficient (or \emph{scalable}) if it requires resources that grow as a polynomial in $n$. But, in practice, a less precise notion of efficiency (scalability) is often more important: the resources required are small for all relevant $n$. The creation of efficient benchmarks is an increasing focus of the benchmarking research community, due to the ongoing rapid increase in the number of qubits in quantum computers. Scalable benchmarks include modern versions of RB \cite{Hines2023-tz, McKay2023-bx, Proctor2022-yl, Hines2023-vq}, cycle benchmarking \cite{Erhard2019-wk}, and mirror circuit benchmarks \cite{Proctor2021-wt, Hines2023-be}. However, there are many widely-used benchmarks (e.g., cross-entropy benchmarking [XEB] \cite{Boixo2018-kp} and the quantum volume benchmark \cite{Cross2019-ku}) that are not scalable, because they rely on classical simulations of general quantum circuits (see Box 4).

\begin{figure*}[t!]
    \centering
    \includegraphics[width=17.8cm]{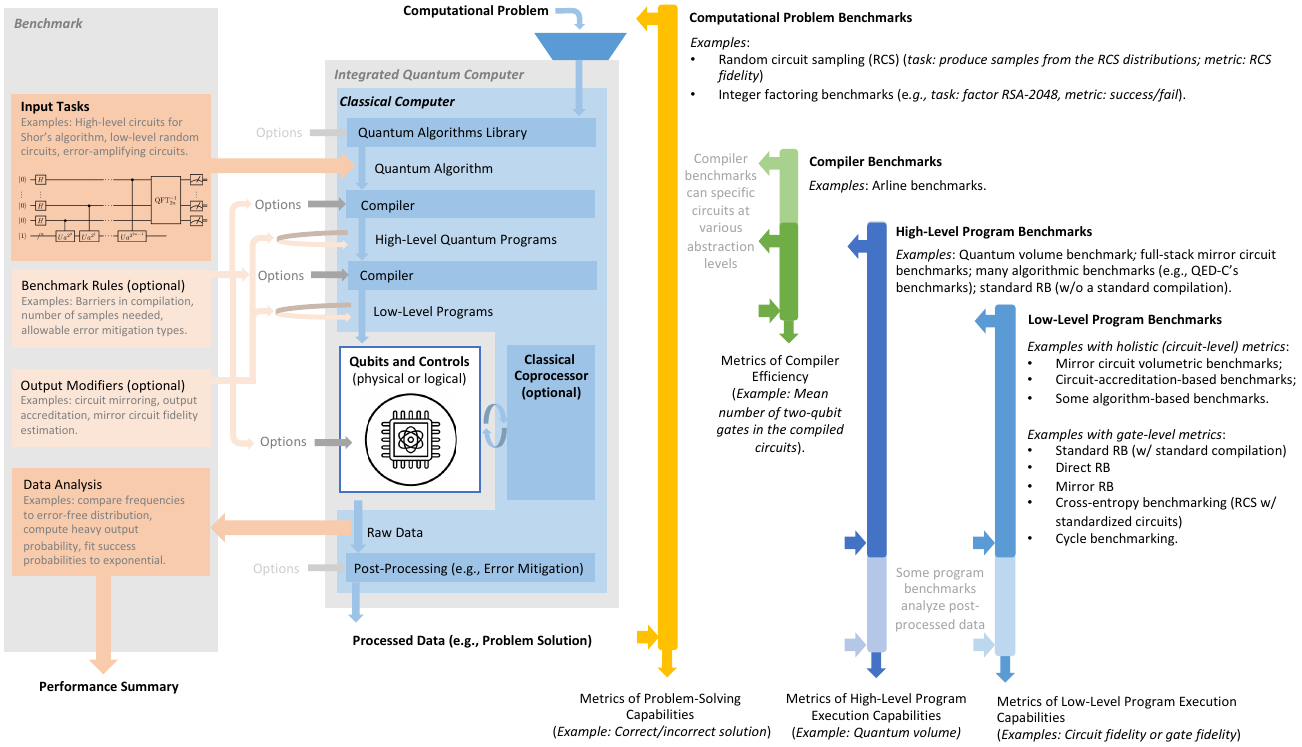}
    \caption{\textbf{How benchmarks interact with integrated quantum computers}. Benchmarks test the joint performance of one or more parts of an integrated quantum computer's ``stack'' (its qubits, compilers, routers, etc). They do so by inserting tasks into one level of the stack, and then analyzing output from the same (or a lower) level of the stack. Benchmarks can limit or adjust what each layer of the stack does (e.g., limiting the types of compilation), which can enable robust and efficient benchmarking. Benchmarks that enter and exit the stack at different levels measure fundamentally different aspects of performance, and form different categories of benchmark. Four important categories are shown here.
    \label{fig:stack}}
\end{figure*}

\section*{Kinds of quantum computer benchmark}
There are many different kinds of quantum computer benchmark, and a bewildering array of specific benchmarks. We think that they can be best categorized and understood by how they interact with integrated quantum computers. Contemporary (prototype) quantum computing systems consist of multiple layers of hardware and software that is often called the quantum computing ``stack'' (see Fig.~\ref{fig:stack}) in analogy to ``software stacks'' in classical computing  \cite{Corcoles2020-vn}. How a benchmark interacts with this stack defines what parts of the quantum computer it tests, what performance metrics it measures, and whether the benchmark measures those metrics effectively. In this section, we explain this framework for understanding benchmarks. We then discuss three kinds of benchmark that are, in our view, of particularly broad importance.

\subsection*{Benchmarks and the quantum computing stack}
Quantum computing systems can be divided into interacting subsystems in many ways, but in our context it is most useful to adopt a division that reflects the flow of information though a computation (Fig.~\ref{fig:stack}). Although different quantum computing systems can have very different architectures, most systems' information flow is approximately as follows. A quantum computing system's input is a computational problem (e.g., ``factor 15'') and its output is classical data that is intended to solve this problem (e.g., $3 \times 5 = 15$). The output is computed by running quantum programs on qubits, wrapped within some potentially complex classical computing systems (that may be automated or human-aided).

A quantum computer's classical computing systems select an algorithm and then realize it as a \emph{quantum program}. Here, a quantum program means a set of quantum circuits (see Box 2) that are perhaps wrapped inside a classical program, as in hybrid algorithms \cite{Preskill2018-jz, Bharti2022-rz, Chen2023-rx}. That quantum program is typically a \emph{high-level quantum program}, i.e., it is written in a high-level language and its quantum circuits contain operations that are not native to that system's computational qubits (which could be physical qubits or logical qubits built out of physical qubits running QEC). For example, it might use circuits that contain two-qubit gates between any pair of qubits (many systems have restricted connectivity), or large subroutines like the $n$-qubit quantum Fourier transform. This high-level program therefore typically needs to be compiled into a \emph{low-level quantum program} whose circuits contain only gates that are ``native'' to that system's computational qubits, before it can be executed. Finally, the system executes that low-level program, processes the data (which can include complex processing like error mitigation \cite{Chen2023-la, Cai2023-sj}) and returns results. This conceptualization of the stack is an idealization, but it is a powerful aid for understanding and designing benchmarks. 

A benchmark should have a well-defined procedure, and this requires that its interactions with the stack are precisely stated. Most existing benchmarks' interactions with the stack fall within the following simple schema shown in Fig.~\ref{fig:stack}. The benchmark inserts computational tasks into one layer of the stack (the \emph{input layer}) and those tasks propagate down the stack until they reach the benchmark's \emph{output layer}. That output is then analyzed to compute performance metrics. For example, the quantum volume benchmark inserts high-level circuits (which the system compiles and runs) and it analyzes the resultant data to estimate the system's quantum volume \cite{Cross2019-ku}. Some benchmarks also specify limitations on what each layer of the stack is allowed to do (\emph{benchmark rules} in Fig.~\ref{fig:stack}), and can even modify the intermediate outputs of the stack (\emph{output modifiers} in Fig.~\ref{fig:stack}). These esoteric aspects of a benchmark can be critical to its robustness or efficiency, e.g., output modifiers can circumvent the problem of efficiently verifying the outputs of general quantum circuits (see Box 4).

A benchmark quantifies the integrated performance of all the computational stages in between (and including) its input and output layers. Benchmarks that enter and exit the stack at different points quantify fundamentally different aspects of performance, and they constitute different and complementary categories of benchmark. Any entry and exit point in the stack defines a category of benchmarks. Some categories of benchmarks test only classical components of a quantum computer, such as compilers  \cite{Kharkov2022-yw, Singh2023-in} or qubit control electronics \cite{Efthymiou2024-qn}. For example, compiler benchmarks \cite{Kharkov2022-yw, Singh2023-in} isolate and test compilation algorithms by inserting high-level quantum circuits into the stack, and then analyzing the low-level circuits the compiler produces (using metrics such as the number of gates in those circuits). However, in our view, the benchmarks of most broad interest and importance test the computational qubits and gates. Three such categories of benchmark are shown in Fig.~\ref{fig:stack}, and we discuss these kinds of benchmark in the remainder of this section.

\subsection*{Benchmarking problem solving capabilities}
A quantum computing system's problem solving capabilities can be directly quantified using \emph{computational problem benchmarks} that challenge it to solve computational problems and quantify performance in terms of metrics for the quality of (and/or time to) the solution. As illustrated in Fig.~\ref{fig:stack}, these benchmarks enter the stack at the highest level, and exit it at the lowest level, i.e., they jointly test an entire computing system. One reason these benchmarks are appealing is that they can be applied to any computing system, enabling comparison between radically different kinds of quantum computers, as well as direct comparisons with classical computers. 

Any computational problem can be used to define a computational problem benchmark, but it is appealing to use problems that are believed to be intractable for classical computers yet efficiently solvable with quantum algorithms. An example is the task ``factor RSA-2048'' with the binary performance metric ``were the correct factors found?'' Benchmarks like this will become increasingly important once quantum utility is first achieved, although note that such benchmarks must address the often-challenging problem of how to quantify the quality of a purported solution to a classically intractable problem (see Box 4).

Computational problem benchmarks that use practical, classically intractable problems provide few insights into the performance of today's quantum computers, because they appear to be far from solving such problems (see Fig~\ref{fig:stack}). This has motivated benchmarking quantum computers' abilities to solve classically tractable problems (e.g., ``factor 15''), but benchmarks of this sort can easily mislead. This is because any such problem can be solved by a quantum computer using primarily (or even only) its classical computing resources. This is illustrated by a series of high-profile experiments that factored 15 using a compilation of Shor’s algorithm that was only possible because 15’s factors were already known \cite{Smolin2013-vc}. Quantum computing hardware capabilities are often demonstrated by running quantum algorithms on simplified problem instances \cite{Google_AI_Quantum_and_Collaborators2020-qw, Harrigan2021-qs, Graham2022-ki}, but these experiments are rarely intended to define formal benchmarks. A good benchmark based on solving classically tractable problems must restrict the kinds of classical computation that a quantum computer is allowed to use when running the benchmark (see Box 3). For this reason, most algorithm-based benchmarks that are used today  \cite{Chen2022-dm, Linke2017-mr, Wright2019-zj, Tomesh2022-nu, Murali2019-my, Donkers2022-wt, Finzgar2022-aa, Mills2020-zh, Lubinski2023-zy, Lubinski2024-ci, Lubinski2023-mr, Chen2023-la, Benedetti2019-pp, Li2020-ry, Quetschlich2023-bg, Dong2021-gj, Martiel2021-vp, Van_der_Schoot2022-gv,Van_der_Schoot2023-vo, Cornelissen2021-yt, Georgopoulos2021-hh, Dong2022-ga} are best categorized as high- or low-level program benchmarks, rather than computational problem benchmarks, which are two categories of benchmark that we now discuss.

\begin{boxfloat}[t!]
\begin{center}
\fbox{%
\hspace{-0.05cm}
\begin{minipage}{.47\textwidth}
\small{
\vspace{0.05cm}
\textbf{Box 3. The role of compilation in benchmarking.} Quantum programs can be written at many levels of abstraction, and the level used in a benchmark's programs fundamentally impacts what it measures. The core of any quantum program is one or more quantum circuits, and a circuit is expressed using gates from some set, also known as a \emph{basis}. This gate set could contain high-level gates (e.g., the $n$-qubit quantum Fourier transform), or only canonical one- and two-qubit gates (e.g., the CNOT and Hadamard gates), or only a particular system's low-level gates (e.g., cross-resonance gates between pairs of connected qubits). However, a benchmark will only be executable on a broad range of quantum computing systems if it permits its circuits to be \emph{compiled} into circuits over different gate sets (we use ``compilation'' to mean the process of turning one circuit into another circuit that implements the same overall computation, which is sometimes subdivided into various interrelated tasks such as gate synthesis, transpilation, and routing). More generally, a quantum program refers to an equivalence class of programs, i.e., ``run this program'' means ``run any program in this program's family'', and a well-defined benchmark whose tasks are quantum programs specifies the equivalence class for each of its programs.

\hspace{0.3cm}
Integrated quantum computers can contain powerful classical computers, and effective, robust benchmarks must take this into account when defining the kinds of compilation that are allowed. This is illustrated by the \emph{de facto} standard RB method \cite{Magesan2011-hc}. $n$-qubit standard RB measures the mean error rate of a set of high-level $n$-qubit quantum circuits, the set of all $n$-qubit Clifford unitaries, that can be compiled in any way. Standard RB measures the ``error-per-Clifford'' by running circuits consisting of $d+1$ random Clifford unitaries (with $d$ varied) that, if composed together, implements the trivial, identity operation. This enables precise estimation of the error-per-Clifford, but only if sequential Clifford unitaries are not compiled together---which is typically described using \emph{compilation barriers} between each of the Clifford unitaries. In our framework of Fig.~\ref{fig:stack}, standard RB is a high-level program benchmark that (like many useful benchmarks) specifies rules that limit what the compiler is allowed to do.

\hspace{0.3cm}
Many high-level program benchmarks are highly permissive in the kinds of compilations they allow, e.g., each of the quantum volume benchmark’s circuits $C$ can be replaced by any circuit $C’$ that, in the absence of errors, implements approximately the same unitary as $C$. This is a common approach, as benchmarks like this jointly test a system's compilation algorithms and qubits. A similar and conceptually simpler approach is for a benchmark to allow its circuits to be replaced with any other circuits that (in the absence of error) produce samples from the same probability distribution, but such benchmarks can be gamed. This is because the benchmark permits replacing each of its circuits $C$ with a potentially trivial circuit $C'$ that simply encodes the already-computed output distribution of $C$, and this is feasible to do (i.e., a system’s classical compilers could find such a $C'$) whenever each $C$ can be quickly simulated classically. 

\vspace{0.05cm}
}\end{minipage}
\hspace{-0.05cm}}
\end{center}
\end{boxfloat}

\subsection*{Benchmarking circuit execution capabilities}
Quantum computations are implemented by running quantum programs, and so many benchmarks directly test a system's program-running capabilities. They typically do so by (i) tasking a quantum computer to run a set of quantum programs, and (ii) computing a metric that quantifies the error in those programs' execution (Box 4 discusses the challenge of \emph{efficiently} computing such metrics). Because the core of any quantum program is the execution of quantum circuits, many of these benchmarks’ tasks are simply quantum circuits rather than more complex programs. Program (or circuit) benchmarks vary in the amount of compilation they allow (see Box 3), i.e., how high up they enter the stack, and they can be broadly categorized as \emph{high-level} or \emph{low-level program benchmarks}.

Program benchmarks also vary in where they exit the stack. Some program benchmarks (such as RB) specify analysis of the raw bit string data from circuits. Some explicitly permit data post-processing including complex error mitigation \cite{Chen2022-dm}. Others are ambiguous about what, if any, post-processing is allowed. Our view is that the results of benchmarks run with error mitigation should currently be interpreted with extreme caution, because clear connections between the error-mitigated performance of today's prototype quantum computers and progress towards quantum utility have not been made.

High-level program benchmarks (see Fig.~\ref{fig:stack}) permit broad optimizations (compiling, routing, etc) of their programs or circuits. The most widely-used such benchmark is the quantum volume benchmark, which quantifies a system's performance on random circuits containing random two-qubit gates coupling arbitrary pairs of qubits \cite{Cross2019-ku}. In contrast, many of the recently-developed high-level program benchmarks are based on applications and algorithms and therefore use highly structured programs \cite{Chen2022-dm, Linke2017-mr, Wright2019-zj, Tomesh2022-nu, Murali2019-my, Donkers2022-wt, Finzgar2022-aa, Mills2020-zh, Lubinski2023-zy, Lubinski2024-ci, Lubinski2023-mr, Chen2023-la, Benedetti2019-pp, Li2020-ry, Quetschlich2023-bg, Dong2021-gj, Martiel2021-vp, Van_der_Schoot2022-gv, Van_der_Schoot2023-vo, Cornelissen2021-yt, Georgopoulos2021-hh, Dong2022-ga}. Prominent examples include the QED-C's benchmarking suite \cite{Lubinski2023-zy,Chen2022-dm} and SuperMarQ \cite{Tomesh2022-nu}. These suites consist of an assortment of benchmarks each of which is based on an algorithm, including algorithms that are illustrative but unlikely to be useful (e.g., the Bernstein–Vazirani algorithm) \cite{Lubinski2023-zy}, heuristic NISQ algorithms \cite{Tomesh2022-nu, Martiel2021-vp}, algorithms that are likely to require a fault-tolerant architecture to provide utility (e.g., Shor's algorithm) \cite{Lubinski2023-zy}, and algorithms that prepare canonical quantum states like GHZ states \cite{Tomesh2022-nu}.

High-level program benchmarks are increasingly being used to compare different quantum computers \cite{Moses2023-do, Lubinski2023-zy, Tomesh2022-nu, Van_der_Schoot2022-gv, Martiel2021-vp, Donkers2022-wt}, and it has been claimed that this enable simple and fair comparisons of systems with very different architectures. In our view, these comparisons can easily mislead. Contemporary quantum computers have not achieved quantum utility, so our perspective is that assessing these systems' program-executing capabilities is useful only in so much as it assesses or incentivizes progress towards the goal of quantum utility. But improved performance on high-level program benchmarks can be obtained via system improvements that are unlikely to bring quantum utility closer. For example, many algorithms are unlikely to provide utility without fault tolerance, so an improvement to a quantum computing system's classical compilation algorithm that enables better NISQ implementations of those algorithms (and therefore improves performance on existing algorithmic benchmarks) does not indicate progress towards quantum utility. The development of high-level program benchmarks that can reliably quantify, incentivize, and compare the progress of disparate architectures (e.g., NISQ versus fault-tolerant architectures) towards computational utility would be extremely valuable for the field.

Low-level program benchmarks (see Fig.~\ref{fig:stack}) forbid intrusive classical compilation of their programs. This prevents improvements on these benchmarks being driven by compilation algorithm optimizations. Examples of low-level program benchmarks include many RB methods \cite{Emerson2005-fd, Emerson2007-am, Knill2008-jf, Magesan2011-hc, Hines2023-tz, McKay2023-bx, Proctor2022-yl, Hines2023-vq, Combes2017-kr, Helsen2019-cp, Helsen2022-yp, Erhard2019-wk}, XEB \cite{Boixo2018-kp}, mirror circuit benchmarks \cite{Proctor2022-yl}, and some algorithm-based benchmarks. Low-level program benchmarks typically permit only \emph{localized} compilation of their circuits, i.e., each individual gate in a circuit can be replaced with a sequence of native operations that synthesize that gate (see Box 3). Low-level program benchmarks directly quantify the performance of a system’s qubits and gates, so they enable discovery and quantification of unexpected errors in qubits and gates, by comparing actual performance to that predicted by a model \cite{Proctor2021-wt, McKay2019-ca, Proctor2022-yl}. Some low-level program benchmarks compute holistic performance metrics, such as \emph{capability regions} \cite{Proctor2021-wt} (see Fig.~\ref{fig:progress}) that directly quantify circuit execution error, but many of them are designed to extract error rates for individual logic operations.

\begin{boxfloat}[t!]
\begin{center}
\fbox{%
\hspace{-0.05cm}
\begin{minipage}{.47\textwidth}
\small{
\vspace{0.05cm}
\textbf{Box 4. The verification problem in benchmarking.} Most benchmarks quantify how well a quantum computer executed its tasks, but designing benchmarks that measure such metrics \emph{efficiently} is difficult in general. Computational problem benchmarks (see Fig.~\ref{fig:stack}) need to quantify the accuracy or correctness of a purported solution to their problem(s). The correctness of the solutions to some classically intractable problems (those in the co-NP complexity class, like factoring) can be easily validated or falsified, but this is atypical of problems with known quantum speedups. In those cases, natural metrics for solution quality will not typically be efficient to measure \cite{Hangleiter2019-bx, Hangleiter2023-jy}, and so a good benchmark must instead measure an efficient-to-estimate and reliable proxy for those metrics. Google's famous random circuit sampling (RCS) experiments utilized a procedure of this sort, specialized to the RCS problem \cite{Arute2019-mk}. More general methods for verifying the outputs of quantum computations have been developed \cite{Hangleiter2019-bx, Hangleiter2023-jy, Carrasco2021-ep, Eisert2020-an, Gheorghiu2019-pa, Hangleiter2024-sz}, including approaches based on cryptographic techniques \cite{Gheorghiu2019-pa} or comparisons to a trusted quantum computer \cite{Elben2020-ru}. However, some of these methods make strong assumptions about the nature of a quantum computer's errors.

\hspace{0.3cm}
Program and circuit benchmarks (see Fig.~\ref{fig:stack}) face a similar verification challenge: many natural metrics for quantifying how well a quantum computer executed a quantum circuit compare the observed and error-free outcome distributions of that circuit \cite{Lubinski2023-zy, Hines2023-be}. However, directly calculating these metrics requires classically computing the error-free outcome distribution, which is exponential expensive in the number of qubits in general. Many benchmarks avoid this problem by using circuits that are efficiently simulable classically \cite{Magesan2011-hc, Hines2023-tz, McKay2023-bx, Proctor2022-yl, Hines2023-vq}. However this approach is typically inappropriate for high-level program benchmarks, because efficiently simulable circuits are often particularly easy to compile into shallow (or even trivial) circuits (see Box 3). One solution to this challenge is to define high-level program benchmarks using circuits that are potentially intractable to simulate classically, and to then ``intercept'' the compiled low-level circuits before they are run (see \emph{output modifiers} in Fig.~\ref{fig:stack}), replacing them with proxies for those circuits whose performance is provably similar but that are efficiently simulable \cite{Hines2023-be, Proctor2022-zs, Ferracin2021-vh}.

\vspace{0.05cm}
}\end{minipage}
\hspace{-0.05cm}}
\end{center}
\end{boxfloat}

\subsection*{Benchmarking components and subroutines} 
Many of the most mature benchmarks are designed to measure the error rates of the fundamental logic operations from which quantum circuits are built (e.g., individual single-qubit and two-qubit gates, layers of gates, measurements, etc). These \emph{component benchmarks} (see Fig.~\ref{fig:kinds-of-benchmark}) typically achieve this by running low-level circuits containing the components of interest and inferring how those components performed from the performance of those circuits. RB protocols \cite{Emerson2005-fd, Emerson2007-am, Knill2008-jf, Magesan2011-hc, Proctor2019-gf, Hines2023-tz, McKay2023-bx, Proctor2022-yl, Hines2023-vq, Combes2017-kr, Helsen2019-cp, Helsen2022-yp, Magesan2012-dz} are the paradigmatic component benchmarks.

RB methods run low-level circuits containing random gates, which average out the details of those gates' errors. This causes the mean success rate of these circuits to decay exponentially in circuit depth, at a rate equal to the mean of those gates' average-case error rate (i.e., process infidelity, or, equivalently, average gate infidelity \cite{Hashim2024-om}). The \emph{de facto} standard RB method measures the error rate of the set of all one- or two-qubit Clifford gates \cite{Magesan2011-hc}. There are now many RB methods and closely-related techniques that adapt or improve standard RB. Prominent examples include direct \cite{Proctor2019-gf}, mirror \cite{Proctor2022-yl, Hines2023-vq}, binary \cite{Hines2023-tz}, character \cite{Helsen2019-cp, Helsen2022-yp}, filtered \cite{Helsen2022-yp, Heinrich2022-cs}, and interleaved RB \cite{Magesan2012-dz}, as well as XEB \cite{Boixo2018-kp} and cycle benchmarking \cite{Erhard2019-wk}. XEB, direct RB, mirror RB, and binary RB are closely related methods for measuring the average error rate of a set of native gate layers. Interleaved RB and cycle benchmarking are methods for estimating the error rate of a \emph{single} gate or layer of gates. Character RB and other filtered RB methods extend RB to sets of gates that are (or generate) groups other than the Clifford group (or another unitary 2-design).

Tomographic characterization protocols \cite{Chuang1997-dj, Gale1968-hp, Hradil1997-tg, Poyatos1997-jv,  Nielsen2021-nu, Hashim2024-om} are sometimes repurposed to benchmark components. The primary purpose of tomography protocols is to estimated detailed, predictive error models. For example, quantum process tomography uses a lot of data to estimate the $4^n \times 4^n$ process matrix describing a gate or subroutine \cite{Chuang1997-dj, Nielsen2021-nu, Hashim2024-om}. A tomographically-estimated model does not constitute a performance metric, but summary performance metrics such as gate fidelity can be extracted from them, and this constitutes a benchmark. This approach is very data-intensive, but once tomography is complete, essentially any metric can be estimated. This includes metrics that no known benchmark can measure efficiently, including worst-case error rates such as diamond norm error \cite{Kliesch2021-ea}. Worst-case metrics are valuable, but are rarely reported because they are pessimistic \emph{and} grow rapidly harder to measure with the number of qubits $n$. Tomography-derived benchmarks are rarely used for more than two qubits.

Benchmarks that can measure the performance of fundamental operations on \emph{logical} qubits will soon become increasingly important. Most of today's benchmarks can be applied to logical qubits \cite{Combes2017-kr} but they were primarily designed for physical qubits. Whether the theories and assumptions underpinning existing benchmarks (e.g., approximate Markovianity \cite{Blume-Kohout2022-ln}) will apply to logical qubits is largely unknown. However, we anticipate that new component benchmarks that are designed for logical qubits will be needed, e.g., because some logical qubit operations (such as lattice surgery \cite{horsman2012surface}) are fundamentally different from any operations on physical qubits.

The move towards fault-tolerant architectures also brings immediate needs for component benchmarks that measure the properties of physical qubits and gates that are most relevant in this setting. Many benchmarks measure the total rate of errors in layers of gates \cite{Proctor2019-gf, Proctor2022-yl, Hines2023-vq, Hines2023-tz, McKay2023-bx, Erhard2019-wk}, which can predict the failure rates of circuits executed directly on physical qubits (e.g., NISQ algorithms). However, in fault-tolerant architectures some kinds of errors are much more damaging than others, because they are more costly to correct  \cite{Campbell2017-tw}. Similarly, some physical operations and subroutines are central to fault-tolerant quantum computing but are not essential in NISQ architectures. Important examples include mid-circuit measurements, parity checks, and syndrome extraction cycles \cite{Bluvstein2023-dp, Krinner2022-tp, Google_Quantum_AI2023-yr, Gupta2024-pr}. Some prototype benchmarks exist for the physical primitives of fault-tolerant quantum computing \cite{Govia2022-ne, Hines2024-qj, Zhang2024-zp, Bluvstein2023-dp, Krinner2022-tp, Google_Quantum_AI2023-yr, Gupta2024-pr}, but mature benchmarks are a near-term need for the field. 

\section*{Measuring progress to quantum utility}
We conclude our perspective by discussing what is, in our view, the most compelling current and near-term purpose of quantum computer benchmarks: measuring progress toward quantum utility. The single biggest challenging here is that ``quantum utility'' is not a uniquely defined goal. There are many important problems that a quantum computer might solve, and many possible routes to creating such a computer.  Our view is that tracking progress towards quantum utility will require a range of complementary benchmarks that are motivated by \emph{challenge problems} (computational problems that, if solved by a quantum computer, would constitute quantum utility), and interpreted relative to \emph{resource estimates} (an accounting of the computational resources required to solve a challenge problem) and \emph{hardware roadmaps} (plans for how to build a quantum computer that can solve a challenge problem). In this section, we discuss challenge problems, resource estimates, and hardware roadmaps---summarizing each area from the perspective of how it impacts benchmarking---and then we conclude by presenting our views on how benchmarks can measure progress towards quantum utility.

\subsection*{Challenge problems for quantum computing}
Quantum computers will not outperform classical computers on all computational tasks, but there's good reason to believe they can speed up specific structured tasks. If such a task is also (subjectively) \emph{useful}, then executing that task successfully can constitute quantum utility.  A challenge problem is a specific instance of a computational problem that is crafted to be feasible to solve with a quantum computer but infeasible---or at least more costly---with any other computer, and is also useful to solve.
 
The most elegant challenge problems involve calculating quantities that cannot feasibly be computed without a quantum computer. Although problems that are just very hard for classical computers are also appealing, they are moving targets. This is because classical algorithms continue to improve and the power and speed of classical computers continues to grow, so even a $1000 \times$ quantum advantage could be erased within years (or even days, with rapid classical algorithm development to counter a claimed quantum advantage)~\cite{beguvsic2024fast,tindall2024efficient,anonymous2024quantum,Fu2024-ub}. So the most compelling challenge problems are specific instances of computational problems for which quantum computers have \emph{exponential advantage} in a key resource (typically time-to-solution).

Two of the best studied categories of challenge problems are integer factorization and quantum chemistry. Instances of factorization are easy to describe (e.g., ``factor this 2048-bit semiprime''~\cite{gidney2021factor, Willsch2023-gq}), they scale naturally in size and difficulty, and their solutions can be verified easily. Instances of quantum chemistry can be stated precisely (e.g., ``estimate an energy eigenvalue of the FeMo cofactor to at least chemical accuracy''~\cite{lee2021even}), but their solutions are typically only heuristically verifiable, and ``hardness'' is challenging to quantify systematically even though a domain expert can usually assess a given instance's classical difficulty.

Few sharply defined challenge problems have been formulated, and the existing challenge problems suggest that achieving quantum utility is very difficult. But there is a huge gray area containing important problems that \emph{might} admit quantum speedups and quantum-friendly problems that \emph{might} be useful to solve. This has fostered confusion about how soon quantum utility might be achieved. Discrete optimization, machine learning, and linear algebra offer promising candidate challenge problems, but it's not clear whether any useful speedups exist in these areas. However, identifying even one additional unambiguous challenge problem from among them (e.g., via the development of a new or improved quantum algorithm) could change the benchmarking landscape drastically.
 
To identify more challenge problems, we may need to rely on heuristic notions of verification and difficulty.  If so, consensus will only emerge from detailed analyses of the best classical algorithms' capabilities and substantive debate between domain experts.  Recent developments in quantum chemistry illustrate this process.  Active debate about what instances are ``hard''~\cite{li2019electronic}, and about which quantum advantages will be both feasible and useful~\cite{lee2023evaluating}, has proceeded in parallel with advances in quantum algorithms for simulation and the careful constant-factor resource estimates~\cite{reiher2017elucidating, lee2021even, von2021quantum} that are necessary to assess quantum utility.
 
 \begin{figure*}
    \centering
    \includegraphics[width=17.5cm]{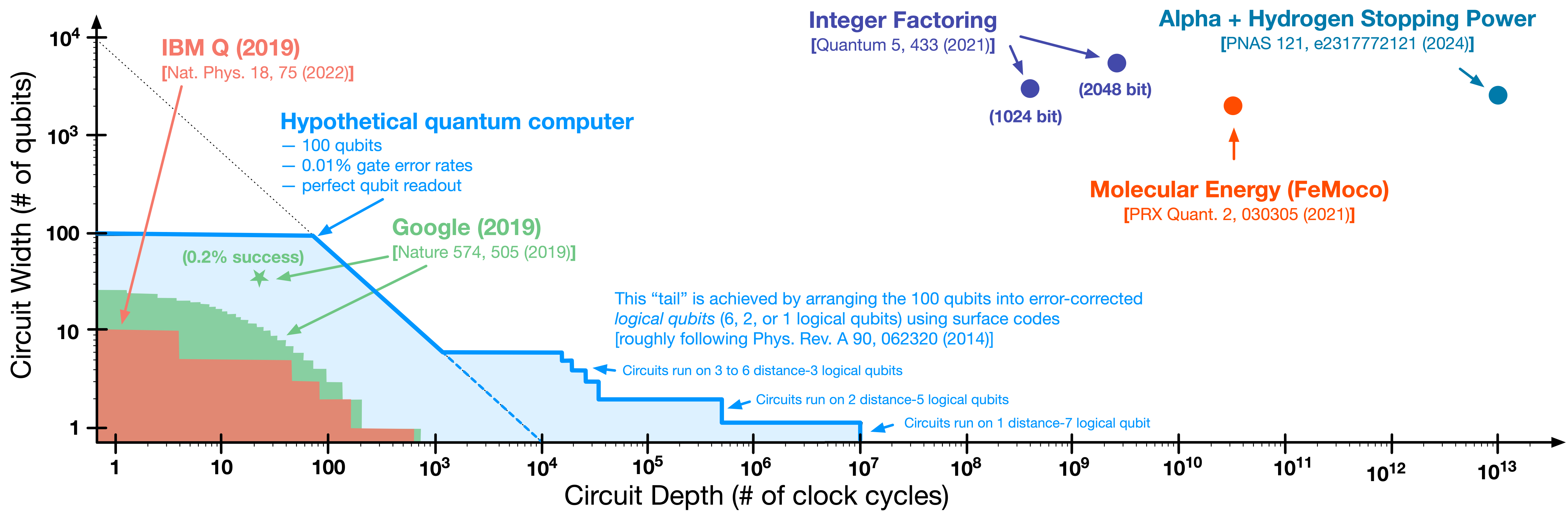}
    \caption{\textbf{Assessing quantum computer performance via capability.} This figure illustrates one way to compare experimentally benchmarked performance against resource estimates for challenge problems, using a multidimensional \emph{capability} metric.  Challenge problems and benchmark tasks are represented by the width (some measure of the number of qubits) and depth (some measure of the number of clock cycles) of a quantum circuit that performs the task. Regions indicate the circuits performable by two real-world quantum computers---Google’s Sycamore (green) as extrapolated from results in Arute \emph{et al.} \cite{Arute2019-mk}, and an ensemble of IBM Q devices (pink) benchmarked by our group \cite{Proctor2021-wt}---and one hypothetical quantum computer (blue) (we use a success threshold of $1/e$). Points indicate constant-factor resource estimates for three candidate challenge problems analyzed in the literature \cite{gidney2021factor, lee2021even, Rubin2024-gc}. For these problems, width is the number of logical qubits, not accounting for logical qubits used in distillation or routing, and depth is the total number of non-Clifford operations (i.e., Toffoli and/or T gates).  These metrics are somewhat crude, but indicate the rough scale of resources required for these challenge problems (but note that the required resources could decrease with new or improved quantum algorithms). We emphasize the wide gulf between that ``utility'' scale and even our hypothetical 100-qubit quantum computer (with beyond state-of-the-art capabilities)---logarithmic axes were required to compress both scales into one figure. Plots like this one could enable stakeholders to track and extrapolate the growth of quantum computer capabilities over time, toward eventual achievement of quantum utility.\label{fig:progress}}
\end{figure*}

\subsection*{Resource estimates for challenge problems}
Resource estimation addresses the question ``How big, fast, and reliable would a quantum computer need to be?''~in order to solve a specific computational problem~\cite{reiher2017elucidating,babbush2018encoding,sanders2020compilation,campbell2021early,gidney2021factor,lee2021even,lemieux2021resource,su2021fault,von2021quantum,delgado2022simulating,goings2022reliably,kim2022fault,berry2023quantum,pathak2023quantifying,rubin2023fault,steudtner2023fault,zini2023quantum,agrawal2024quantifying,bellonzi2024feasibility,berry2024analyzing,clinton2024towards,cortes2024fault,elenewski2024prospects,fomichev2024simulating,nguyen2024quantum,otten2024quantum,penuel2024feasibility,pathak2024requirements,rhodes2024exponential,saadatmand2024fault,Rubin2024-gc}. A resource estimate for a challenge problem is a description of a minimal (known) quantum computer that would suffice to solve that problem.  We say ``minimal'' rather than ``minimum'' because some resources can be traded for others (e.g., memory vs speed, or space vs time).

Quantum algorithms are typically first analyzed in terms of their asymptotic scaling, i.e., how the required computational resources scale with problem size, solution accuracy, etc. Resource estimation means that the constant prefactors for specific problem instances are also calculated. Resource estimates are quantitative, and indicate whether theoretical quantum speedups are practically feasible.  For example, recent resource estimates \cite{babbush2021focus, hoefler2023disentangling} provide a strong argument that good challenge problems are unlikely to involve merely quadratic quantum speedups, even under optimistic assumptions about quantum hardware advances and pessimistic assumptions about the performance of classical hardware.

Resource estimates can be specified at various levels of abstraction and detail, and different levels of detail could motivate the use or development of entirely different benchmarks. A simple \emph{logical resource estimate} for a challenge problem might be ``$N$ qubits, capable of executing [specific gates], with error probability $\leq\epsilon$ per gate, running for time $T$,'' derived directly from the quantum circuits of an algorithm known to solve the problem. Logical resource estimates for most existing challenge problems mandate error probabilities of $\epsilon < 10^{-15}$/gate, which are widely believed to be achievable only via fault-tolerant QEC.  

More granular resource estimates account for the unavoidable existence of ``hard'' and ``easy'' logic operations in every fault-tolerant architecture (see Box 2). They count the minimum number and cost of ``hard'' operations that will need to be implemented using tricks like magic-state distillation and injection~\cite{litinski2019magic} or code switching~\cite{kubica2015universal}.  A more detailed approach specifies a particular QEC code and method of implementing fault-tolerant computation within that code (e.g., lattice surgery~\cite{horsman2012surface}). The most sophisticated \textit{physical resource estimates} provide a full architectural specification---``$N'$ \emph{physical} qubits, implementing [specific fault-tolerant quantum computing scheme], with error probability $\leq\epsilon'$ per \emph{physical} gate, running for time $T'$.'' Even the best existing physical resource estimates are based on oversimplified error models (e.g., assuming each gate's errors are fully described by a single error rate), and so they do not set the goalposts for quantum utility in terms of detailed and realistic hardware specifications.

\subsection*{Roadmaps to solving challenge problems}
Resource estimates for challenge problems tell us where the goalposts of quantum utility are. \emph{Roadmaps} specify an engineering route to get to those goalposts---they specify how a particular quantum computing technology can plausibly progress towards quantum utility. A roadmap describes a sequence of increasingly capable prototype devices. Some steps in this sequence may introduce key technologies that don't immediately increase computational power and/or aren't rewarded by today's most widely-used (computational) benchmarks. Roadmaps enable a broader view of benchmarking that recognizes and quantifies all the steps along a path to utility.
 
Each prototype in a roadmap can be associated with a list of measurable specifications, e.g., gate error rates, latency times, crosstalk rates, control bandwidth, etc. A roadmap is (at minimum) a sequence of such specification lists, which trace out a path to a useful device.  Real devices are enormously complex, with many specifications.  To track progress along a roadmap using benchmarks, a short list of key specifications needs to be identified.  Those specifications should be measurable, and sufficient to prove that the prototype achieves its goals within the context of the roadmap.
 
The most relevant specifications (and thus the benchmarks necessary to infer them) will likely change at different points along a roadmap. For example, in the early stages of a roadmap toward a fault-tolerant quantum computer with millions of qubits \cite{gidney2021factor, lee2021even, Rubin2024-gc}, gate error rates on physical qubits may be paramount and require direct benchmarking. In later stages, after multiple fault-tolerant logical qubits have been assembled, \textit{logical} qubit error rates may replace them as a key specification. A robust understanding of what the most important specifications are, at each stage of a particular quantum computing technology's development, will be important for the development of benchmarks that successfully track progress towards utility.
 
\subsection*{Benchmarks that measure progress towards utility}
In the context provided by challenge problems, resource estimates, and a roadmap, 
we can assess whether a particular benchmark measures progress along that roadmap, and (if necessary) design new benchmarks that do measure this progress. An as-built prototype can be compared to an idealized prototype in a specific roadmap (see Fig.~\ref{fig:schematic}b) by using benchmarks that directly measure each of the key performance specifications of that prototype. This enables answering questions such as ``are crosstalk error rates as low as planned?''. The benchmarks needed for this task will vary with the scale of the prototype and the details of the roadmap. 
As quantum computing technology progresses, the most relevant specifications for a prototype quantum computer will be increasingly formulated in terms of high-level metrics (e.g., logical qubit error rates, or circuit success probabilities). One way to measure metrics like this is to do so \emph{indirectly} by measuring the properties of the lowest-level components (e.g., error rates of physical one- and two-qubit gates) and combining them using calculations or simulations to infer the values of higher-level metrics (e.g., logical qubit error rates). This approach is implicit in fault-tolerance threshold theorems that compute high-level computational properties from low-level models. However, quantum computers are complex systems, whose emergent behavior may not be reliably captured or predicted by modeling informed by component benchmarks \cite{Proctor2021-wt, Hines2023-vq, McKay2019-ca, Proctor2022-yl}, so this indirect approach to benchmarking is unlikely to be sufficient.

Benchmarks that measure performance metrics that are specific to a particular roadmap will be essential for informing the engineering of useful quantum computers. But there is also a need to quantify progress towards utility with high-level, intuitive, and technology-agnostic performance metrics. Holistic benchmarks that probe \emph{computational} power directly can provide intuitive and direct evidence of progress along a roadmap. It is unlikely that any single scalar metric of computational power will accurately assess or incentivize progress toward quantum utility. However, holistic benchmarks can report rich multi-dimensional metrics with greater descriptive power. A particular example that we have found useful and inspiring is \emph{capability benchmarking} \cite{Proctor2021-wt}.

Since a quantum computer's computational power comes from running quantum programs, one way to probe that power is to ask ``What quantum programs can it run?'' The set of quantum programs that a computer can run, with reasonable accuracy using reasonable time and energy, is called its \textit{capability} \cite{Proctor2021-wt, Hothem2023-dq}. Formally, this set-valued metric is infeasible to measure because there are far too many quantum programs. But it can be sketched or approximated using a short list of key \emph{program features}. If sufficiently faithful, a sketch of capability provides intuitive heuristic answers to many questions about a quantum processor's computational ability.

Capabilities can be sketched by choosing a limited set of program features, such as circuit width and circuit depth \cite{Blume-Kohout2020-de, Proctor2021-wt} or the number of ``hard'' gates in a circuit \cite{Chen2023-la}. We define a \emph{circuit class} containing all programs with the same values of the selected features.  Ideally, we seek to choose features so that a given processor will be able to successfully run every circuit in a class, or none of them. Inasmuch as this holds, a benchmark of moderate complexity can probe the processor's ability to run sample programs from each circuit class.  The result is a high-dimensional metric that can be visualized as a \textit{capability
region} \cite{Proctor2021-wt} (Fig. \ref{fig:progress}) in the space defined by the program features. These capability regions can be compared directly to resource estimates, for challenge problems, displayed on the same plot, as well as computed for future prototypes on a roadmap.

This specific approach has clear limitations that have not yet been overcome (What kind of programs should be run?  Do good features exist?  How many are needed? What defines ``the same'' program on different architectures? Should the programs be defined at high or low levels of abstraction? What kinds of compilation should be allowed?). But we suspect that any successful strategy for benchmarking progress to utility will need to move beyond the single-scalar-metric paradigm, and will need to address these challenges or similar ones.

An open problem for any such approach is how to find or construct genuinely representative \textit{proxy programs} that behave like programs that would solve a challenge problem, but that can be scaled seamlessly to fit on any given prototype. A related open question is how these programs change when adapted to different architectures, and how to construct benchmarks that evaluate both NISQ and fault-tolerant architectures on equal footing. There might be a significant era during which advanced NISQ computers compete with, and should be fairly compared to, early fault-tolerant quantum computers. They may pursue very different challenge problems via very different roadmaps. Creating benchmarks that can compare progress of \textit{all} architectures in a shared context seems challenging, but valuable to many stakeholders.

\bibliography{bibliography}

\section*{Acknowledgements}
This material was funded in part by the U.S. Department of Energy, Office of Science, Office of Advanced Scientific Computing Research, Quantum Testbed Pathfinder Program. T.P. acknowledges support from an Office of Advanced Scientific Computing Research Early Career Award. A.D.B. acknowledges support from the National Nuclear Security Administration’s Advanced Simulation and Computing Program and the Department of Energy (DOE) Office of Fusion Energy Sciences “Foundations for quantum simulation of warm dense matter” project. Sandia National Laboratories is a multi-program laboratory managed and operated by National Technology and Engineering Solutions of Sandia, LLC., a wholly owned subsidiary of Honeywell International, Inc., for the U.S. Department of Energy's National Nuclear Security Administration under contract DE-NA-0003525. All statements of fact, opinion or conclusions contained herein are those of the authors and should not be construed as representing the official views or policies of the U.S. Department of Energy, or the U.S. Government.

\section*{Author contributions}
All authors contributed to developing the perspective presented here. TP and RBK led the writing of the manuscript, with all authors contributing. 
\section*{Competing interests}
The authors declare no competing interests.
\end{document}